\documentstyle[prb,aps]{revtex}

\begin{document}
\title{FLOWER-LIKE SQUEEZING IN\ THE\ MOTION OF A LASER-DRIVEN TRAPPED\ ION}
\author{Nguyen Ba An$^{1,2}$\thanks{%
Corresponding author. Email: nbaan@netnam.org.vn} and Truong Minh Duc$^{3}$}
\address{$^{1}$Institute of Physics, P.O.Box 429 Bo Ho, Hanoi 10000, Vietnam\\
$^{2}$Faculty of Technology, Vietnam National University, 144 Xuan Thuy, Cau%
\\
Giay, Hanoi, Vietnam\\
$^{3}$Physics Department, Hue University, 32 Le Loi, Hue, Vietnam}
\maketitle
\date{}

\begin{abstract}
We investigate the $Nth$ order amplitude squeezing in the fan-state $\left|
\xi ;2k,f\right\rangle _{F}$ which is a linear superposition of the $2k$%
-quantum nonlinear coherent states. Unlike in usual states where an ellipse
is the symbol of squeezing, a $4k$-winged flower results in the fan state.
We first derive the analytical expression of squeezing for arbitrary $k,$ $%
N, $ $f$ and then study in detail the case of a laser-driven trapped ion
characterized by a specific form of the nonlinear function $f.$ We show that
the lowest order in which squeezing may appear and the number of directions
along which the amplitude may be squeezed depend only on $k$ whereas the
precise directions of squeezing are determined also by the other physical
parameters involved. Finally, we present a scheme to produce such fan-states.

\end{abstract}


\noindent {\bf 1. Introduction}

\noindent Due to quantum interference a superposition state may have
properties qualitatively different from those of its component states (see,
e.g., Refs. \onlinecite{1,2,3,4,5,6,7,8}). For example, while coherent
states are most classical states, their linear superpositions may exhibit
various nonclassical behaviors. Inversely, an infinite superposition of Fock
states with absolutely uncertain phases may result in a state with a
definite certainty in phase.

The study of squeezed state has continuously progressed since its first
introduction in 1970 \cite{stoler} because this state promises potential
applications in communication networks, detecting extremely weak fields,
waveguide tap, etc. (see, e.g., Ref. \onlinecite{wall}) and teleportation of
entangled states \cite{gorbachev}. Generalizations of squeezed states
towards higher-order \cite{HM,MH1} and multi-mode \cite{MH2,kumar,antinh}
situations have also been made. Usually in an amplitude-squeezed state one
quadrature phase reduces the quantum noise level below the standard quantum
limit, i.e. squeezing occurs in one direction. Since the amplitude component
along the squeezing direction could be exploited for practical applications
more squeezing directions would be of physical relevance. In fact, in a
recent paper \cite{an1} multi-directional amplitude squeezing has been shown
possible in the so called fan-state which is superposed from a finite number
of multi-quantum nonlinear coherent states each of which is capable of
possessing squeezing only in a single direction.

The fan state is defined as (see Ref. \onlinecite{an1} and the references
therein for more details) 
\begin{equation}
\left| \xi ;2k,f\right\rangle _{F}=D_{k}^{-1/2}\sum_{q=0}^{2k-1}\left| \xi
_{q};2k,f\right\rangle  \label{fan}
\end{equation}
where $k=1,2,3,...;$ $\xi _{q}=\xi \exp (i\pi q/2k)$ with $\xi $ a complex
number; 
\begin{equation}
D_{k}=D_{k}(|\xi |^{2})=\sum_{m=0}^{\infty }\frac{\left| \xi \right|
^{4km}\left| J_{k}(m)\right| ^{2}}{(2km)!\left[ f(2km)(!)^{2k}\right] ^{2}},
\label{c}
\end{equation}
with $J_{k}(m)=\sum_{q=0}^{2k-1}\exp (i\pi qm),$ comes from the
normalization condition and 
\begin{equation}
\left| \xi _{q};2k,f\right\rangle =\sum_{n=0}^{\infty }\frac{\xi _{q}^{2kn}}{%
\sqrt{(2kn)!}f(2kn)(!)^{2k}}\left| 2kn\right\rangle ,  \label{8}
\end{equation}
with $\left| 2kn\right\rangle $ a Fock state, is a sub-state of the more
general so-called multi-quantum nonlinear coherent states \cite
{kncs1,kncs2,kncs3}, the eigenstates of the operator $a^{K}f(\widehat{n})$
with $K=2,3,4,...$ and $f$ an arbitrary real nonlinear operator-valued
function of $\widehat{n}=a^{+}a$ with $a$ the bosonic annihilation operator.
In Eqs. (\ref{c}) and (\ref{8}) the notation $(!)^{2k}$ is understood as
follows 
\begin{equation}
f(p)(!)^{2k}=\left\{ 
\begin{array}{lll}
f(p)f(p-2k)f(p-4k)...f(q) & \text{if} & p\geq 2k,\text{ }0\leq q<2k \\ 
1 & \text{if} & 0\leq p<2k
\end{array}
\right. .
\end{equation}
In the complex plane the $\xi _{q}$ in Eq. (\ref{fan}) are oriented like an
open paper fan (see Fig. 1). Hence the state $\left| \xi ;2k,f\right\rangle
_{F}$ is referred to as fan-state identified by the subscript $``F"$
implying ``fan''. Since for $k=1$ the fan shrinks to a setsquare, the state 
\begin{equation}
\left| \xi ;2,f\right\rangle _{F}=D_{1}^{-1/2}\left( \left| \xi
;2,f\right\rangle +\left| i\xi ;2,f\right\rangle \right)  \label{f2}
\end{equation}
was named orthogonal-even nonlinear coherent state \cite{das} which is the
simplest fan state. In particular, when $f\equiv 1$ the state (\ref{f2})
reduces to that proposed in Ref. \onlinecite{lynch}. In Ref. \onlinecite{an1}
the $N^{th}$ order amplitude squeezing was directly calculated several first
values of $k$ and $N$ (namely, $k=1,2$ and $N=2,4,6,8)$ using $f\equiv 1,$
showing explicitly the multi-directional character of squeezing. Since the
presence of squeezing in more than one direction would provide more choices
to precisely determine the quantum state of a field for possible
applications, a further more general and more realistic study of
multi-directional squeezing proves desirable and necessary. In Section 2 we
shall derive for the higher-order squeezing the general expression valid for
arbitrary $k,$ $N$ and $f.$ In Section 3 we shall consider a specific
situation associated with the vibrational motion of the center-of-mass of a
laser-driven trapped ion for which the function $f$ and the quantity $\xi $
are specified by \cite{kncs2,kncs3} 
\begin{equation}
f\left( \widehat{n}+K\right) =\frac{\widehat{n}!L_{\widehat{n}}^{K}(\eta
^{2})}{\left( \widehat{n}+K\right) !L_{\widehat{n}}^{0}(\eta ^{2})},\text{ }%
\xi ^{K}=-\frac{\text{e}^{i\varphi }\Omega _{0}}{(i\eta )^{K}\Omega _{1}}
\label{fxi}
\end{equation}
where $L_{n}^{m}(x)$ is the $n^{th}$ generalized Laguerre polynomial in $x$
for parameter $m,$ $\eta $ is the Lamb-Dicke parameter, $\varphi =\varphi
_{1}-\varphi _{0}$ with $\varphi _{0}$ ($\varphi _{1}$) the phase of the
driving laser which is resonant with (detuned to the $K^{th}$ lower sideband
of) the electronic transition of the ion, and $\Omega _{0,1}$ the
corresponding pure electronic transition Rabi frequencies. Besides $K,$
there are two more physical parameters in Eqs. (\ref{fxi}): the $\eta $
which is controllable by the trapping potential and the $\xi $ which is
controllable by the driving laser fields. The motivation of expanding the
results in Ref. \onlinecite{an1} for $f\equiv 1$ to the case of trapped ions
for $f$ given by (\ref{fxi}) is that trapped ions can be used to implement
quantum logic gates \cite{lg1,lg2} and so far various kinds of nonclassical
states have been proposed and observed \cite{nc1,nc2,nc3,nc4,nc5} in the
motion of a trapped ion. In Section 4 a scheme to produce the fan-state will
be presented and the final section is the Conclusion.

\noindent {\bf 2. General expression for the higher-order amplitude squeezing%
}

\noindent Let a field amplitude component pointing along the direction
making an angle $\Phi $ with the real axis in the complex plane be 
\begin{equation}
X_{\Phi }=\frac{1}{\sqrt{2}}\left( a\mbox{e}^{-i\Phi }+a^{+}\mbox{e}^{i\Phi
}\right)  \label{am}
\end{equation}
where the operators $a,$ $a^{+}$ obey the commutation relation $[a,a^{+}]=1.$
According to Ref. \onlinecite{HM}, a state $\left| ...\right\rangle $ is
said to be amplitude-squeezed to the $N^{th}$ order ($N$ an even integer)
along the direction $\Phi $ if 
\begin{equation}
\left\langle (\Delta X_{\Phi })^{N}\right\rangle <R_{N}=\frac{\left(
N-1\right) !!}{2^{N/2}}  \label{Sn}
\end{equation}
where $\Delta X_{\Phi }\equiv X_{\Phi }-\left\langle X_{\Phi }\right\rangle $
and $R_{N}=\left\langle (\Delta X_{\Phi })^{N}\right\rangle _{CS}$ with the
subscript ``CS'' standing for coherent state. Likewise, the parameter $%
S_{N}(\Phi )=\left\langle (\Delta X_{\Phi })^{N}\right\rangle -R_{N}$ can be
introduced and squeezing occurs whenever $-R_{N}\leq S_{N}(\Phi )<0.$ We
choose the real axis along the direction of $\xi $ allowing to treat $\xi $
as a real number. In the fan state, we obtain 
\[
\left\langle a^{+l}a^{m}\right\rangle _{k}=\frac{\xi ^{(l-m)}}{D_{k}(\xi
^{2})}I\left( \frac{l-m}{2k}\right) 
\]
\begin{equation}
\times \sum_{n=0}^{\infty }\frac{\theta (2kn-m)\xi ^{4kn}J_{k}\left( n+\frac{%
l-m}{2k}\right) J_{k}\left( n\right) }{(2kn-m)!f(2kn)(!)^{2k}f\left(
2kn+l-m\right) (!)^{2k}}  \label{alam}
\end{equation}
where $\left\langle ...\right\rangle _{k}\equiv $ $_{F}\left\langle \xi
;2k,f\right| ...\left| \xi ;2k,f\right\rangle _{F}.$ The function $I(x)$
equals unity if $x$ is an integer and, it is zero otherwise. The presence of
this function is specifically associated with the fan-state $\left| \xi
;2k,f\right\rangle _{F}$ which is a multi-quantum state $(2k=2,4,6,...).$ To
get rid of the step function $\theta (2kn-m)$ we can simply remove it and
replace in the summation $n=0$ by $n=n_{\min }$ with $n_{\min }$ equal to
the integer part of $(m+2k-1)/2k.$ Calculations will be greatly facilitated
by observing the following property of $J_{k}\left( n\right) ,$ 
\begin{equation}
J_{k}\left( n\right) =\left\{ 
\begin{array}{ll}
2k, & n\mbox{ even integers} \\ 
0, & n\mbox{ odd integers}
\end{array}
\right.  \label{J}
\end{equation}
and 
\begin{equation}
J_{k}\left( n\right) J_{k}\left( n+n^{\prime }\right) =\left\{ 
\begin{array}{ll}
2k^{2}\left( 1+(-1)^{n}\right) , & n^{\prime }\mbox{ even integers} \\ 
0, & n^{\prime }\mbox{ odd integers}
\end{array}
\right. .  \label{JJ}
\end{equation}

Noting that in the fan state $\left\langle a\right\rangle _{k}=\left\langle
a^{+}\right\rangle _{k}=0$ and using the Campbell-Baker-Hausdorff identity
we get $\left\langle (\Delta X_{\Phi })^{N}\right\rangle _{k}$ in the form 
\[
\left\langle (\Delta X_{\Phi })^{N}\right\rangle _{k}=\frac{N!}{2^{N}} 
\]
\begin{equation}
\times \sum_{m,l=0}^{\infty }\sqrt{2^{m+l}}\theta (N-l-m)\frac{\left\langle
a^{+m}a^{l}\mbox{e}^{i(m-l)\Phi }\right\rangle _{k}}{m!l!\left(
(N-l-m)/2\right) !}.  \label{d1}
\end{equation}
The double sum over $m$ and $l$ in the r.h.s. of Eq. (\ref{d1}) can be split
into three parts as $\sum_{m,l}x_{ml}\equiv
\sum_{m}x_{mm}+\sum_{m,l>m}x_{ml}+\sum_{m,l<m}x_{ml}.$ In the first sum the $%
m=0$ term equals nothing else but $R_{N},$ while the two last sums are
complex conjugate to each other. We then arrive from Eq. (\ref{d1}) for the
squeeze parameter at 
\[
S_{N}^{(k)}(\Phi )=\frac{N!}{2^{N}}\left\{ \sum_{m=1}^{\infty }2^{m}\theta
(N-2m)\frac{\left\langle a^{+m}a^{m}\right\rangle _{k}}{(m!)^{2}\left(
N/2-m\right) !}\right. 
\]
\begin{equation}
\left. +2\sum_{l>m,m=0}^{\infty }\sqrt{2^{m+l}}\theta (N-l-m)\frac{\cos
[(l-m)\Phi ]\left\langle a^{+m}a^{l}\right\rangle _{k}}{m!l!\left(
(N-l-m)/2\right) !}\right\} .  \label{d2}
\end{equation}
Because of Eqs. (\ref{alam}) and (\ref{JJ}), in Eq. (\ref{d2}) only terms
with $l=m+4pk,$ where $p=1,2,...,P,$ contribute. Dictated by the step
function, the possible maximal value of $p,$ i.e. $P,$ is equal to the
integer part of $N/4k$ and, for a fixed $p,$ the $m$ varies from $0$ to $M$
with $M$ the integer part of $(N/2-2pk).$ As a consequence, Eq. (\ref{d2})
becomes 
\begin{equation}
S_{N}^{(k)}(\Phi )=A_{N}^{(k)}+\sum_{p=1}^{P}B_{N}^{(k)}(p)\cos (4pk\Phi )
\label{S}
\end{equation}
with 
\begin{equation}
A_{N}^{(k)}=\frac{N!}{2^{N}}\sum_{m=1}^{N/2}\frac{2^{m}\left\langle
a^{+m}a^{m}\right\rangle _{k}}{(m!)^{2}\left( N/2-m\right) !}  \label{A}
\end{equation}
and 
\begin{equation}
B_{N}^{(k)}(p)=\frac{4^{pk}N!}{2^{N-1}}\sum_{m=0}^{M}\frac{2^{m}\left\langle
a^{+m}a^{m+4pk}\right\rangle _{k}}{m!(m+4pk)!(N/2-m-2pk)!}.  \label{B}
\end{equation}
In passing we note from Eq. (\ref{alam}) that $\left\langle
a^{+m}a^{m}\right\rangle _{k}$ are always positive but $\left\langle
a^{+m}a^{m+4pk}\right\rangle _{k}$ are not. As a result of these, $%
A_{N}^{(k)}$ are always positive but $B_{N}^{(k)}(p)$ are not.

\noindent {\bf 3. Laser-driven trapped ion}

\noindent The formulas derived in the preceding section are applicable to
arbitrary $k,$ $N$ and $f.$ In this section we deal with a laser-driven
trapped ion. Then $a$ $(a^{+})$ denotes the annihilation (creation) operator
for a quantum of the quantized field of vibrational motion of the
center-of-mass of the ion. The function $f$ takes the form (\ref{fxi})
involving two physically controllable parameters $\eta $ and $\xi .$

From Eq. (\ref{S}) it is clear that, for a given $k,$ the lowest possible
order of squeezing is $N=N_{\min }=4k.$ For $N<4k,$ i.e. $P<1,$ the second
term in Eq. (\ref{S}) does not appear, the first term is $\Phi $-independent
and always positive due to Eq. (\ref{alam}), implying no squeezing. For $%
N\geq 4k,$ i.e. $P\geq 1,$ the $\Phi $-dependence comes into play through
the second term in Eq. (\ref{S}) and may give rise to but is not sufficient
for squeezing. This conclusion is independent of $f.$

It seems impossible in general to establish the sufficient condition for
squeezing. This can be done only in the case of $f\equiv 1$ by an asymptotic
analysis for $S_{N}^{(k)}(\Phi )$ in the limit $\xi \rightarrow 0$ making
use of Eq. (\ref{alam}). Keeping only the leading order in $\xi $ in $%
A_{N}^{(k)}$ and $B_{N}^{(k)}(p),$ we have 
\begin{equation}
S_{N}^{(k)}(\Phi )=\frac{N!}{2^{N-1}}\left[ U+W\cos \left( 4k\Phi \right)
\right]  \label{UW}
\end{equation}
with 
\begin{equation}
U=\sum_{m=1}^{N/2}\frac{2^{m}\theta (4k-m)}{(m!)^{2}(N/2-m)!(4k-m)!}\xi ^{8k}
\label{U}
\end{equation}
and 
\begin{equation}
W=\frac{2^{2k+1}}{(4k)!(N/2-2k)!}\xi ^{4k}.  \label{W}
\end{equation}
As $U$ as well as $W$ are positive and $U<W$ always holds for $|\xi |\ll 1$
there is some $\Phi $ for which $S_{N}^{(k)}(\Phi )<0.$ Hence, for $f\equiv
1,$ squeezing always occurs in the small $|\xi |$ limit (in any order $N\geq
4k,$ of course). This fact has been confirmed in \cite{an1} by direct
analytic calculations. For the specific function $f$ identified by Eqs. (\ref
{fxi}) numerical evaluations should be carried out. The simulation shows
that $B_{N}^{(k)}\equiv B_{N}^{(k)}(1)\gg B_{N}^{(k)}(p)$ with $p\geq 2$ for
any $\eta $ and $\xi ,$ yielding to a very good approximation 
\begin{equation}
S_{N}^{(k)}(\Phi )=A_{N}^{(k)}+B_{N}^{(k)}\cos (4k\Phi ).  \label{Sf}
\end{equation}
Obviously, squeezing arises whenever $\left| B_{N}^{(k)}\right| >A_{N}^{(k)}$
and the squeezing direction is dictated by the sign of $B_{N}^{(k)}.$
Generally, for arbitrary $k$ and $N,$ the $B_{N}^{(k)}$ may be either
positive or negative depending on the parameters $\xi $ and $\eta .$ These
will be elucidated in what follows. Let us define squeezing (stretching)
directions as those along which an amplitude component is maximally squeezed
(stretched). For $k=1$ and $N=4$ the simulation indicates that $B_{4}^{(1)}$
is always positive so that the stretching direction is surely along $\Phi =0$
and the most probable direction for squeezing is along $\Phi =\pi /4.$
Figure 2 plots $S_{4}^{(1)}$ for $\Phi =\pi /4$ as a function of $\xi ^{2}$
and $\eta ^{2}.$ Clearly, $S_{4}^{(1)}<0,$ i.e. squeezing exists, in some
range of the values of $\xi ^{2}$ and $\eta ^{2}.$ The parameter domain in
which squeezing appears is depicted as a phase diagram in the $(\xi
^{2},\eta ^{2})$-plane in Fig. 3. The squeezing region is bounded inside the
solid curve while maximal squeezing takes place along the dashed curve. In
this case $(k=1,N=4)$ the squeezing (stretching) directions are determined
in the same way as for $f\equiv 1$ (see Ref. \onlinecite{an1}), i.e.
squeezing is simultaneously and equally maximal at 
\begin{equation}
\Phi =\Phi _{n}^{(sq)}=\frac{(1+2n)\pi }{4}\mbox{ with }n=0,1  \label{psq}
\end{equation}
while maximal stretching occurs simultaneously and equally at 
\begin{equation}
\Phi =\Phi _{n}^{(st)}=\frac{n\pi }{2}\mbox{ with }n=0,1.  \label{pst}
\end{equation}
In Fig. 4 we display the uncertainty area, that is the polar plot of $%
\left\langle (\Delta X_{\Phi })^{N}\right\rangle _{k},$ for $k=1,$ $N=4$
with the parameters corresponding to point 1 (long-dashed), point 2 (solid)
and point 3 (short-dashed) in Fig. 3. The circle of radius $3/4$ represents
the uncertainty in the respective coherent state. The situation changes
curiously for higher values of $k$ and $N.$ For illustration, let $k=3$ and $%
N=12.$ In Fig. 5 we plot $A_{12}^{(3)},$ $B_{12}^{(3)}$ and $\left|
B_{12}^{(3)}\right| $ for $\xi ^{2}=0.1$ as functions of $\eta ^{2}.$ The
curves intersect at three points $\eta _{1}^{2}\simeq 0.107,$ $\eta
_{2}^{2}\simeq 0.223$ and $\eta _{3}^{2}\simeq 0.367.$ Transparently,
squeezing exists for $\eta $ such that (i) $\eta _{1}^{2}<\eta ^{2}<\eta
_{2}^{2}$ and (ii) $\eta _{2}^{2}<\eta ^{2}<\eta _{3}^{2}.$ The interesting
issue is however that the squeezing directions differ in the two
above-listed situations. Namely, in the situation (i) $B_{12}^{(3)}>0$ and
squeezing is simultaneously and equally maximal at 
\begin{equation}
\Phi =\Phi _{n}^{(sq)}=\frac{(1+2n)\pi }{12}\mbox{ with }n=0,1,...,5
\end{equation}
whereas maximal stretching occurs simultaneously and equally at 
\begin{equation}
\Phi =\Phi _{n}^{(st)}=\frac{n\pi }{6}\mbox{ with }n=0,1,...,5.
\end{equation}
Yet, in the situation (ii) $B_{12}^{(3)}<0$ and there is an exchange in
directions for squeezing and stretching, i.e. squeezing is simultaneously
and equally maximal at 
\begin{equation}
\Phi =\Phi _{n}^{(sq)}=\frac{n\pi }{6}\mbox{ with }n=0,1,...,5
\end{equation}
whereas maximal stretching occurs simultaneously and equally at 
\begin{equation}
\Phi =\Phi _{n}^{(st)}=\frac{(1+2n)\pi }{12}\mbox{ with }n=0,1,...,5.
\end{equation}
Such a directional exchange is demonstrated in Fig. 6 in which polar plots
of $S_{12}^{(3)}$ are shown for $\xi ^{2}=0.1$ while a) $\eta ^{2}=0.2$ and
b) $\eta ^{2}=0.25.$ In Fig. 6 the center corresponds to the coherent state,
whereas squeezing (stretching) shows up as the shorter (longer) wings.

\noindent {\bf 4. Production scheme}

\noindent We now address the issue of how to produce the fan-state defined
and studied in the preceding sections. At that aim we substitute (\ref{8})
into the r.h.s. of (\ref{fan}) and get 
\begin{equation}
\left| \xi ;2k,f\right\rangle _{F}=D_{k}^{-1/2}\sum_{n=0}^{\infty }\frac{\xi
^{2kn}J_{k}(n)}{\sqrt{(2kn)!}f(2kn)(!)^{2k}}\left| 2kn\right\rangle
\label{aa}
\end{equation}
which upon use of (\ref{J}) becomes 
\begin{equation}
\left| \xi ;2k,f\right\rangle _{F}=2kD_{k}^{-1/2}\sum_{n=0}^{\infty }\frac{%
\xi ^{4kn}}{\sqrt{(4kn)!}f(4kn)(!)^{4k}}\left| 4kn\right\rangle .  \label{bb}
\end{equation}
It is a simple matter by using (\ref{c}) to verify that 
\begin{equation}
2kD_{k}^{-1/2}=\left[ \sum_{m=0}^{\infty }\frac{|\xi |^{8km}}{(4km)!\left(
f(4km)(!)^{4k}\right) ^{2}}\right] ^{-1/2}  \label{dd}
\end{equation}
and, thus, the r.h.s. of (\ref{bb}) is nothing else but the normalized state 
$\left| \xi ;4k,0,f\right\rangle $ which is a right eigenstate of the
operator $a^{4k}f(a^{+}a),$%
\begin{equation}
a^{4k}f(a^{+}a)\left| \xi ;4k,j,f\right\rangle =\xi ^{4k}\left| \xi
;4k,j,f\right\rangle ,  \label{cc}
\end{equation}
corresponding to $j=0$ (see Refs. \onlinecite{kncs1,kncs2,kncs3}. As has
been shown in Refs. \onlinecite{kncs2,kncs3}, the state $\left| \xi
;4k,j,f\right\rangle $ can be generated in the well-resolved sideband regime
as the stable steady state solution of the laser-ion system by driving the
trapped ion with two laser beams: the first laser is tuned to be resonant
with the ion transition frequency but the second one is detuned to the ion's
($4k)^{th}$ lower sideband. If initially the ion is cooled down to
zero-point energy \cite{zero} then the state $\left| \xi
;4k,0,f\right\rangle $ is achieved which is the wanted fan-state $\left| \xi
;2k,f\right\rangle _{F}$, as noted above.

\noindent {\bf 5. Conclusion}

\noindent We have studied the $N^{th}$ order amplitude squeezing in the fan
state specified by $k=1,$ $2,$ $3,$ $....$ for a concrete form of the
nonlinear function $f$ characterizing the multi-quantum nonlinear coherent
state of vibrational motion of the center-of-mass of a trapped ion properly
driven by laser fields. The lowest order $N_{\min }$ in which squeezing may
occur depends on $k.$ Namely, $N_{\min }=4k.$ The new physics associated
with the fan state is the possibility of simultaneous squeezing in more than
one direction. Given $k,$ squeezing in any allowed orders, if it exists,
appears simultaneously along $2k$ directions. The number of stretching is
also $2k.$ Each of the $2k$ stretching directions exactly bisects the angle
between two neighboring squeezing directions forming therefore an
uncertainty region in the form of a symmetric $4k$-winged flower (see Fig. 4
for example). By this reason squeezing in the fan state may be referred to
as flower-like squeezing to distinguish it from the usual squeezing for
which the uncertainty area gets the form of an ellipse. For a laser-driven
trapped ion with the specific form of the nonlinear function $f$ the precise
directions of squeezing and stretching are determined also by the physical
parameters involved, i.e. by $\eta $ (the Lamb-Dicke parameter) and $\xi $
(the eigenvalue of the multi-quantum nonlinear coherent state). For $\eta $
and $\xi $ such that $B_{N}^{(k)}>0$ squeezing occurs along $2k$ directions
determined by 
\begin{equation}
\Phi =\Phi _{n}^{(sq)}=\frac{(1+2n)\pi }{4k}\mbox{ with }n=0,1,...,2k-1
\end{equation}
while the stretching directions are determined by 
\begin{equation}
\Phi =\Phi _{n}^{(st)}=\frac{n\pi }{2k}\mbox{ with }n=0,1,...,2k-1.
\end{equation}
On the other hand, if $\eta $ and $\xi $ are such that $B_{N}^{(k)}<0$ the
squeezing and stretching directions are exchanged, i.e. 
\begin{equation}
\Phi _{n}^{(sq)}=\frac{n\pi }{2k}\mbox{ with }n=0,1,...,2k-1
\end{equation}
and 
\begin{equation}
\Phi _{n}^{(st)}=\frac{(1+2n)\pi }{4k}\mbox{ with }n=0,1,...,2k-1.
\end{equation}

\noindent {\bf Acknowledgments}

\noindent This work was supported by a National Basic Research Program
KT-04.1.2 and by the Faculty of Technology of VNU.

\begin{center}
{\bf FIGURE\ CAPTIONS}
\end{center}

\begin{enumerate}
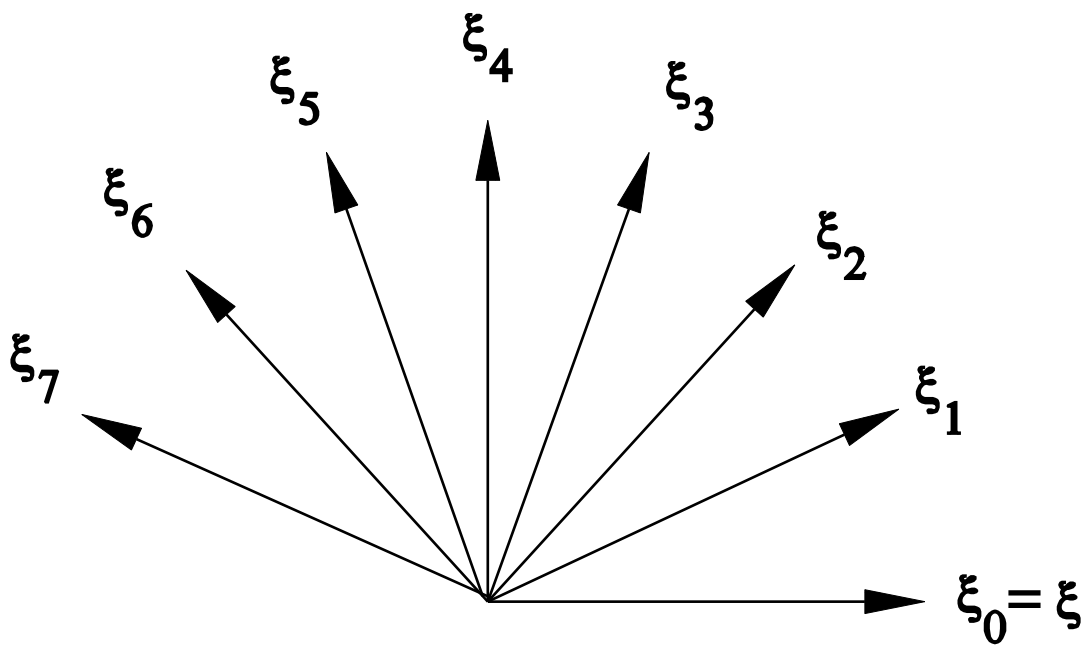
\item[Fig. 1.]  The open fan formed by the $\xi _{q}$ of the fan state $%
\left| \xi ;2k,f\right\rangle _{F},$ Eq. (\ref{fan}), in the complex plane.

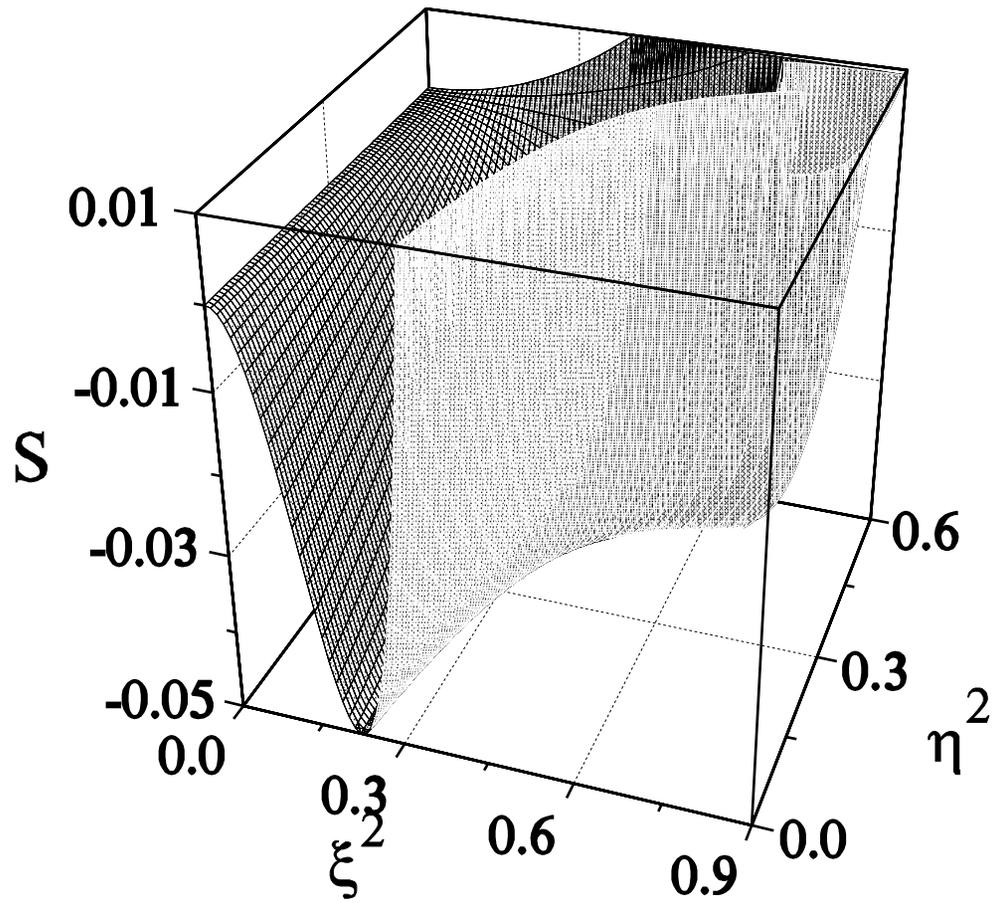
\item[Fig. 2.]  $S\equiv S_{N}^{(k)}(\Phi )$ for $k=1,$ $N=4$ and $\Phi =\pi
/4$ as a function of $\xi ^{2}$ and $\eta ^{2}.$

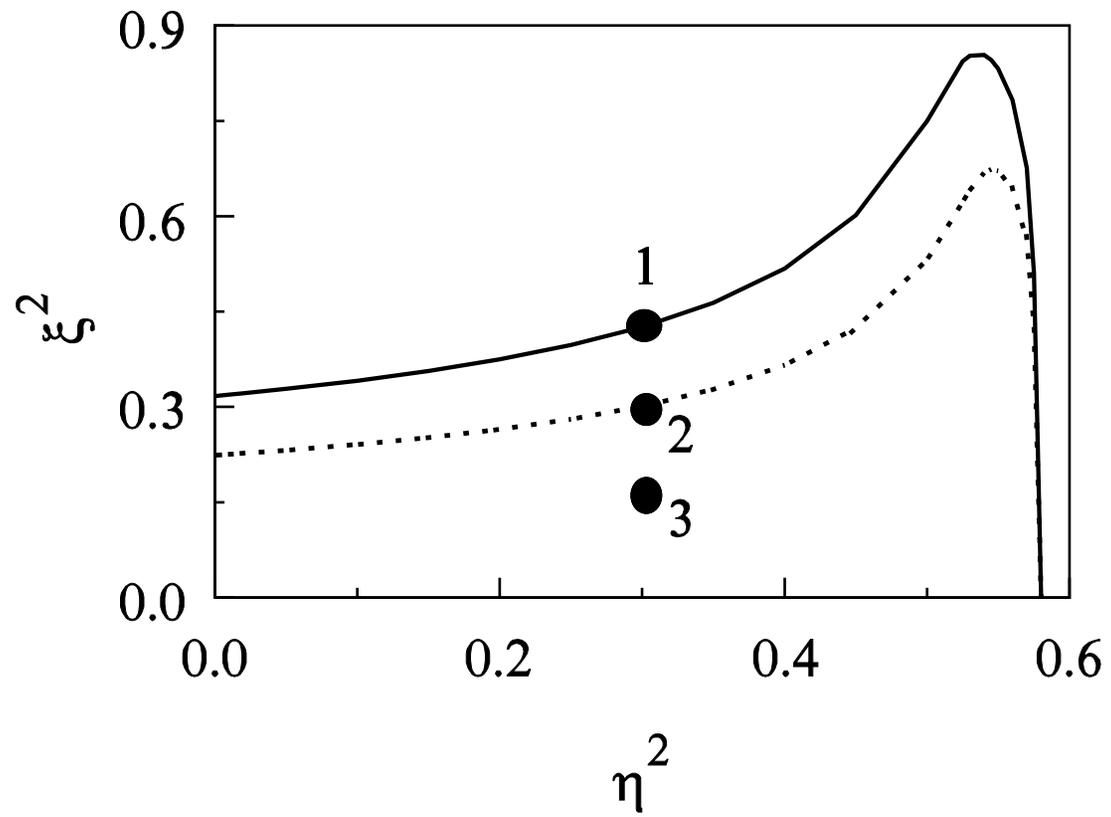
\item[Fig. 3.]  Squeezing phase diagram in the ($\xi ^{2},\eta ^{2})$-plane
for $k=1,$ $N=4$ and $\Phi =\pi /4.$ Squeezing exists inside the domain
bounded by the solid curve. Maximal squeezing occurs along the dashed curve.

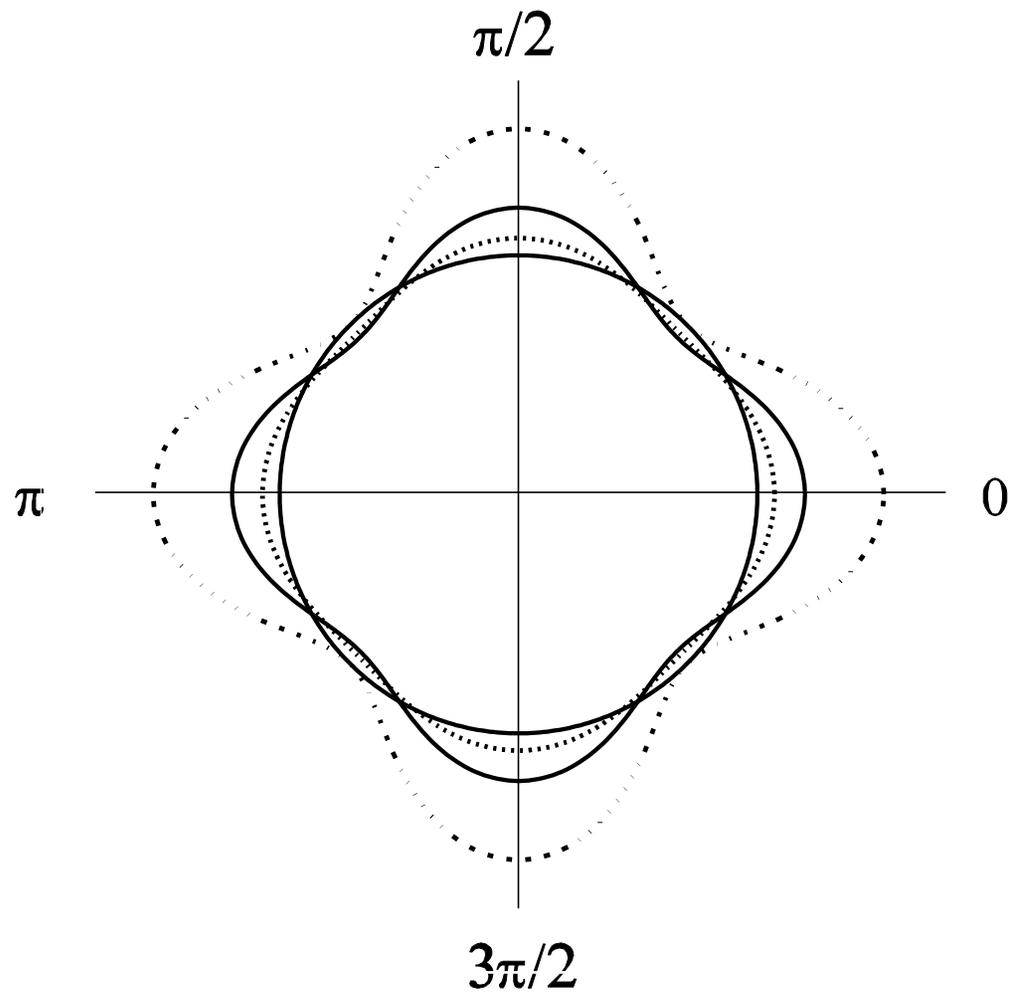
\item[Fig. 4.]  The uncertainty area, i.e. the polar plot of $\left\langle
(\Delta X_{\Phi })^{N}\right\rangle _{k},$ for $k=1$ and $N=4$ with the
parameters corresponding to point 1 (long-dashed), point 2 (solid) and point
3 (short-dashed) in Fig. 3. The circle of radius 3/4 is for the respective
coherent state.

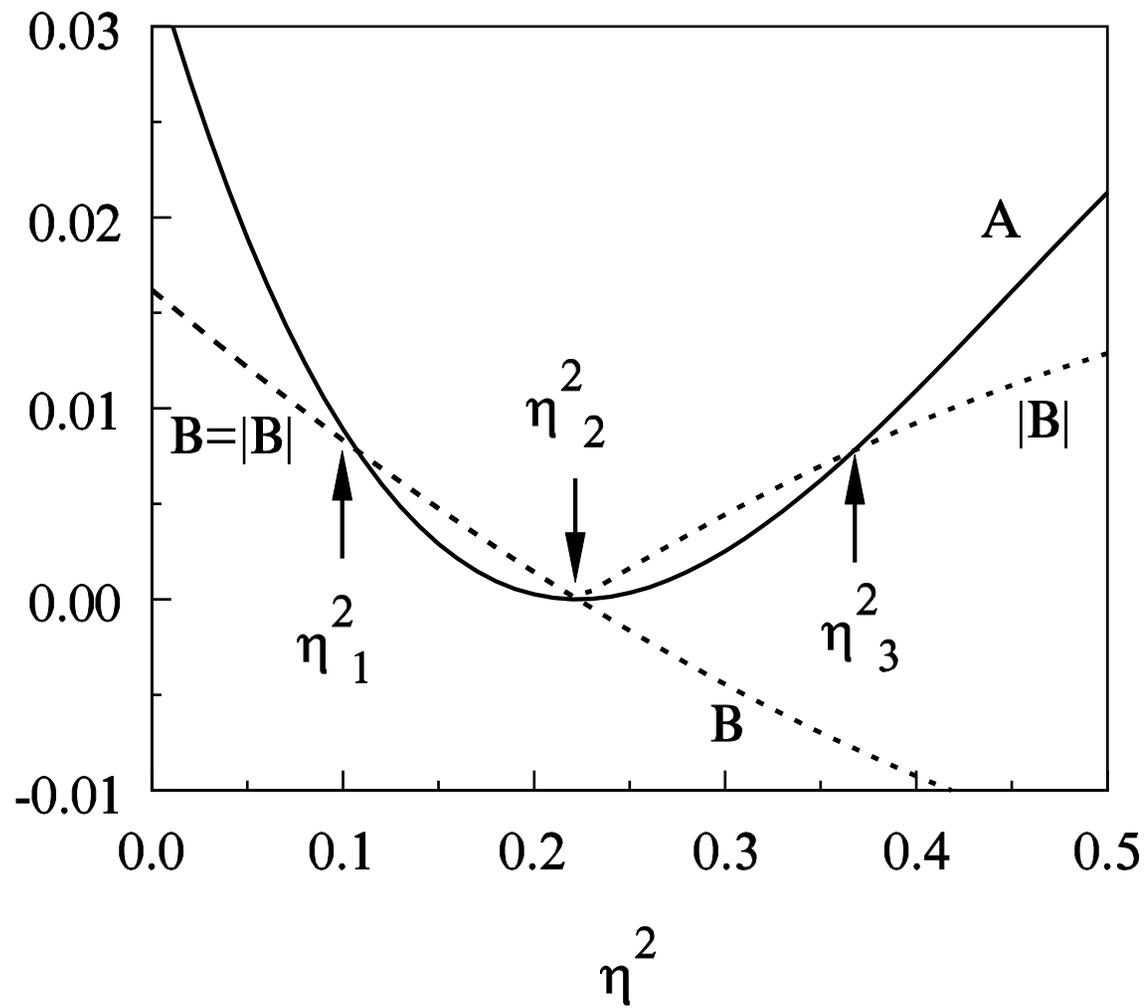
\item[Fig. 5.]  $A_{N}^{(k)}$ (solid curve), $B_{N}^{(k)}$ and $\left|
B_{N}^{(k)}\right| $ (dashed curves) versus $\eta ^{2}$ for $\xi ^{2}=0.1,$ $%
k=3$ and $N=12.$ The curves intersect at $\eta _{1}^{2}\simeq 0.107,$ $\eta
_{2}^{2}\simeq 0.223$ and $\eta _{3}^{2}\simeq 0.367.$

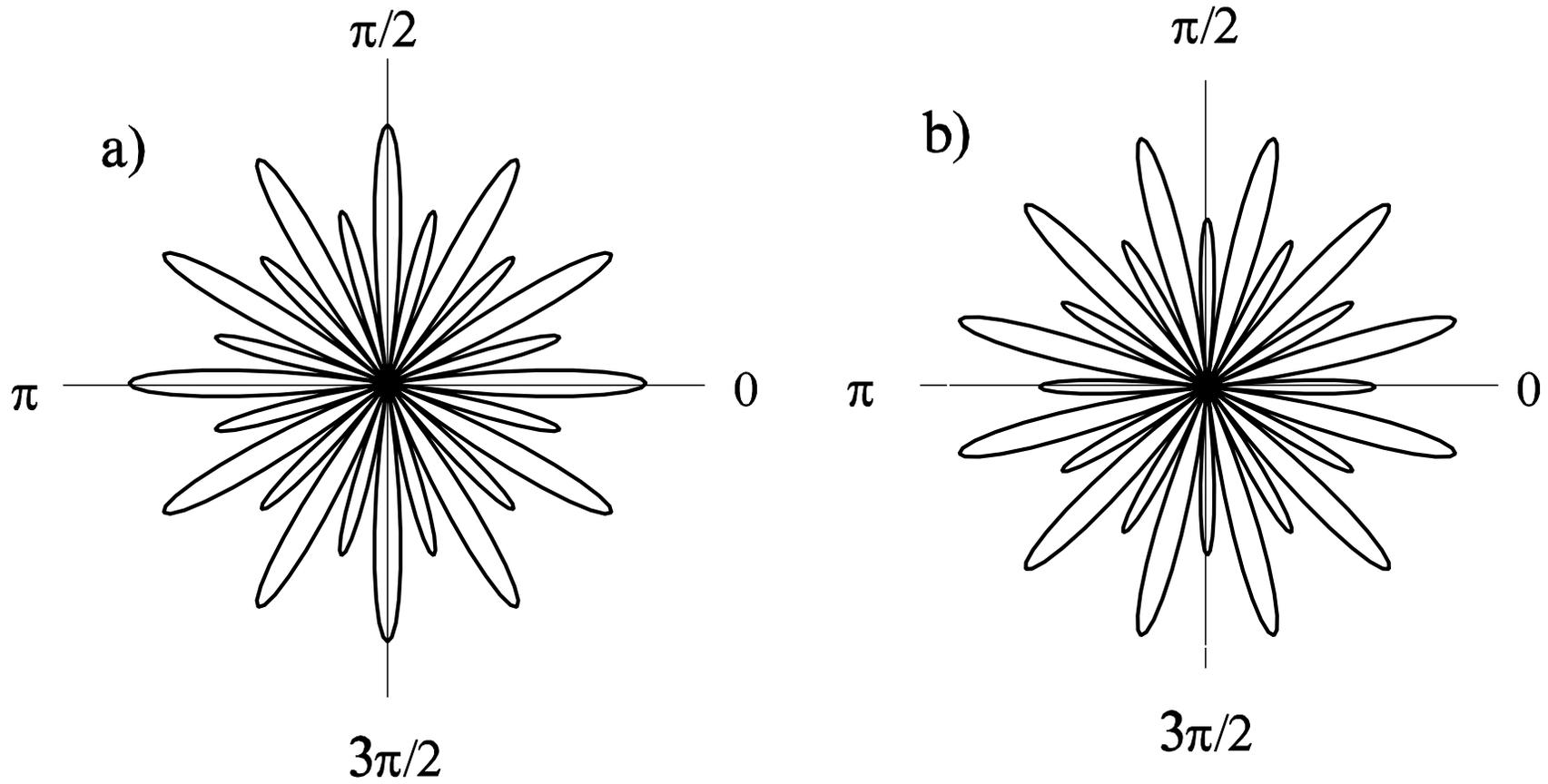
\item[Fig. 6.]  Polar plots of $S_{N}^{(k)}(\Phi )$ for $\xi ^{2}=0.1,$ $%
k=3, $ $N=12$ and a) $\eta ^{2}=0.2:$ squeezing occurs along $\Phi =\Phi
_{n}^{(sq)}=(1+2n)\pi /12$ with $n=0,1,...,5$ ; b) $\eta ^{2}=0.25:$
squeezing occurs along $\Phi =\Phi _{n}^{(sq)}=n\pi /6$ with $n=0,1,...,5.$
\end{enumerate}

\end{document}